\documentclass[runningheads]{llncs}
\usepackage{xcolor}
\usepackage{hyperref}
\usepackage{graphicx}

\begin{document}
\title{Towards Immediate Feedback for Security Relevant Code in Development Environments\thanks{This work is funded by the BMBF project CRITICALMATE (16KIS0995).}}
\titlerunning{Towards Immediate Feedback for Security Relevant Code}
\author{Markus Haug\orcidID{0000-0001-9377-0677} \and
Ana Cristina Franco da Silva\orcidID{0000-0001-8549-350X} \and
Stefan Wagner\orcidID{0000-0002-5256-8429}}
\authorrunning{M. Haug et al.}
\institute{
Institute of Software Engineering, University of Stuttgart, \\Universitätsstraße 38, 70569 Stuttgart, Germany}
\maketitle              %
\begin{abstract}
Nowadays, the correct use of cryptography libraries is essential to ensure the necessary information security in different kinds of applications.
A common practice in software development is the use of static application security testing (SAST) tools to analyze code regarding security vulnerabilities. Most of these tools are designed to run separately from development environments. Their results are extensive lists of security notifications, which software developers have to inspect manually in a time-consuming follow-up step.
To support developers in their tasks of developing secure code, we present an approach for providing them with continuous immediate feedback of SAST tools in integrated development environments (IDEs). 
Our approach also considers the understandability of security notifications and aims for a user-centered approach that leverages developers' feedback to build an adaptive system tailored to each individual developer.

\end{abstract}
\section{Introduction}
An essential practice for ensuring IT security in software applications nowadays is the use of cryptography libraries, such as the Bouncy Castle Crypto APIs\footnote{\url{https://bouncycastle.org}} or the Java security APIs.
For example, in the communication among distributed services in cloud environments, the security of information in transit is a highly important requirement~\cite{iankoulova2012cloud}.
Yet, using cryptography libraries is not easy because of their complexity. They are often used incorrectly, for example, due to misuse of parameters, which can lead to severe security issues~\cite{nadi2016jumping}.

Already during implementation, static application security testing (SAST) tools, such as SonarQube\footnote{\url{https://www.sonarqube.org}} or CogniCrypt~\cite{kruger2017cognicrypt}, can be used to analyze the code in development regarding security vulnerabilities~\cite{2020Nguyen}. 
For this, SAST tools contain comprehensive rule sets, which define, for example, correct or incorrect API usage patterns, and are used to recognize issues and decide when to show security notifications.

Many current SAST tools are designed to be used separately from integrated development environments (IDEs).
That is, security analyses are usually performed as additional steps in build processes, for example, in nightly builds.
This normally leads to extensive lists of security notifications, which software developers have to inspect manually in a follow-up step.
In addition to being time-consuming and cumbersome, the interruption and associated loss of context increase the challenge of understanding and fixing an issue for developers.

In this paper, we present an approach that supports software developers in their tasks of developing secure code.
Using our approach, developers are continuously provided with immediate feedback from SAST tools in their IDEs. 
Our approach considers the understandability of security notifications and also leverages developer feedback to build an adaptive system tailored to each individual developer.
This approach has been developed as a prototype within the BMBF funded project \emph{Cybersecurity static analysis with immediate feedback} (CRITICALMATE\footnote{\url{https://www.forschung-it-sicherheit-kommunikationssysteme.de/projekte/criticalmate}}). 

The remainder of this paper is structured as follows: 
in Section~\ref{sec:approach}, we present our approach.
Section~\ref{sec:related-work} discusses related work. 
Finally, Section~\ref{sec:conclusion} concludes our paper and gives an outlook on future work.

\section{Approach}\label{sec:approach}
Our approach consists of two major parts regarding notifications of security tools: (i) their efficient integration into developer's IDEs to provide immediate feedback, and (ii) the understandability of security notifications.

\subsection{Adaptive Immediate Feedback}\label{sec:immediate-feedback}

To efficiently integrate notifications of SAST tools into development environments, we propose an 
adaptive IDE plugin, which continuously receives analysis results of a SAST tool and shows them in the IDE's code editor as security notifications. 
This approach follows the client-server pattern, where the server is any SAST tool conducting the security analyses, and the client is the IDE plugin receiving the results of the analyses.

An advantage of immediate feedback in IDEs is that the typical workflow interruption while running static code analysis separately is avoided~\cite{smith2020}.
A further advantage of such a tool support is that security issues in the code are recognized as early as possible.
This allows developers to fix issues while they still have most of the context available.

In CRITICALMATE, we have integrated a novel static code analysis engine into IntelliJ, Eclipse, and Visual Studio Code through plugins.
Software developers get immediate feedback about security issues, directly in the code editor of their IDEs.
Any information security issues in the code are underlined in red.
An on-demand pop-up provides additional details about each issue.
Furthermore, in-code annotations are supported to suppress notifications as desired.
This is useful to suppress notifications that developers classify as false positives.

The analysis engine and IDE plugins have been implemented as prototypes and successfully provide immediate feedback.
In the following, we discuss potential improvements to increase their usefulness for developers.

One opportunity for improvement is the false positive rate.
In SAST tools, the increased analysis performance required for immediate feedback frequently comes at the cost of an elevated false positive rate~\cite{2022alahmadi,aloraini2019empirical}.
Generally, such tools have two options to handle a high false positive rate:
(i) they implement more accurate analysis strategies, usually at a performance cost, or (ii) they tighten their confidence threshold. As a consequence, they issue fewer notifications, which however can increase their false negative rate.
Due to the severe consequences of undetected vulnerabilities, we want to ensure a low false negative rate. That is, tightening the confidence threshold is not a suitable option.

Furthermore, a high false positive rate reduces user trust in such tools, which might cause adverse effects.
For example, developers might disable analyses across their whole projects instead of single notifications because of frustrating high false positive rate.
In doing so, they might also inadvertently disable notifications indicating actual security issues in the code.

To handle false positives, we consider an adaptive analysis approach with incremental accuracy.
For each class of issues, the analysis could choose a suitable strategy depending on factors, such as  time budget, user-defined confidence rating, or recorded false positive rate for a certain  combination of issue and strategy.
For example, for a class of notifications known to have a low false positive rate, the analysis could choose a less detailed strategy, which would save time for other analyses.
In contrast, for a class with a low confidence level, the analysis could choose a more accurate iterative strategy, until it achieves a desired confidence level, as long as the time budget permits this.

To be tailored to software developers, the proposed IDE plugin should be able to learn from software developers' feedback, especially about how they assess notifications as false positives in the context of their development projects.
Based on this, the plugin should be able to adapt the classes and amount of shown notifications.
For example, the plugin could learn to suppress notifications, which developers frequently classify as false positives.

Finally, the plugin could also learn the developer's typical workflow, such as if they usually fix the code as soon as notifications are shown or just before committing or sharing their code.
Based on this, the plugin could increase the amount of shown notifications when a developer is most likely to react to them.
At other times, the shown amount would be reduced, e.g., by disabling highlights in the editor and listing the notifications only in the IDE's error list.
This would allow developers to perform their tasks in the way that is most suitable for them.
Reducing the amount of information when developers are unlikely to need it, could also mitigate information overload.

\subsection{Understandable Security Notifications}

Several studies in the literature have shown that many usability issues exist in current SAST tools with respect to understandability of security notifications. 
Tahaei et al.~\cite{tahaei2021security} conducted an experiment aiming to understand how helpful static analysis tools notifications (SonarQube\footnote{\url{https://www.sonarqube.org}} and SpotBugs\footnote{\url{https://spotbugs.github.io}}) are to developers.
They emphasize as a finding that developers make mistakes even if these were well-known mistakes and had known solutions, which suggests a lack of developer awareness and/or missing support in addressing security issues.

Smith et al.~\cite{smith2020} conducted a usability evaluation of four SAST tools.
Within the list of found usability issues\footnote{\url{https://figshare.com/s/71d97832ae3b04e0ff1a}}, there are issues related to the content of the notification, such as “Fix suggestions not adequately explained / sometimes missing” or “Verbose XML output distracting”. 

\begin{figure}[t!]
    \centering
    \includegraphics[width=\textwidth]{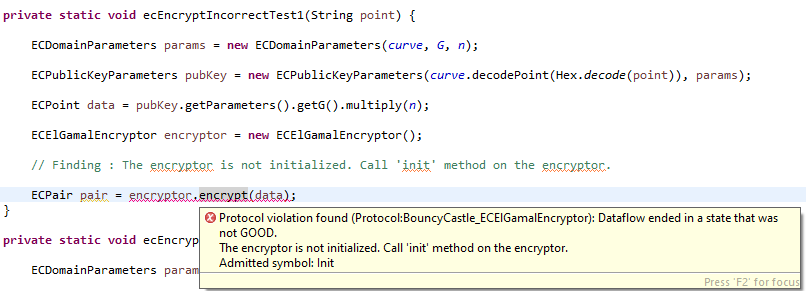}
    \caption{Security notification regarding misuse of Bouncy Castle Crypto API for elliptic-curve (EC) encryption}
    \label{fig:notification_eclipse}
\end{figure}

Figure~\ref{fig:notification_eclipse} shows an example code snippet employing the Bouncy Castle Crypto API for asymmetric ElGamal encryption with an elliptic curve (EC).
The encryptor needs to be initialized by calling the \texttt{init()} method with suitable parameters, i.e., the public key of the recipient, before encrypting the data.
However, the developer forgot to call the \texttt{init()} method in this snippet.

In Figure~\ref{fig:notification_eclipse}, we can also see the IDE notification that the CRITICALMATE analysis engine generates for this example.
While the notification correctly identifies the problem, it is difficult to understand, especially for developers who are not well-versed in the field of cryptography.
Furthermore, there is no example of how to fix the issue, which has been identified as a problem for understandability above.
Consequently, the solution for the issue might not be clear enough because of missing information.

\begin{figure}[t!]
    \centering
    \includegraphics[width=0.8\textwidth]{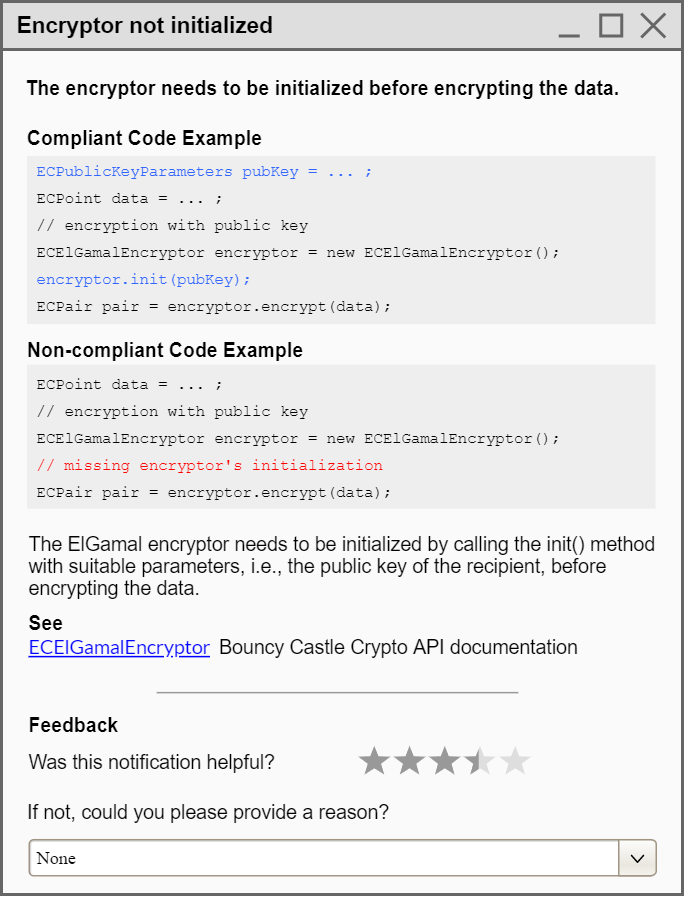}
    \caption{Notification format using as example an API misuse}
    \label{fig:notification-format}
\end{figure}

Therefore, to support software developers in avoiding misinterpretation of security notifications, we propose an approach for presenting security notifications within IDEs that follows usability guidelines to increase the understandability of the security notifications.

Figure~\ref{fig:notification-format} shows our proposed notification format based on the format used in the SonarSource rule repository\footnote{\url{https://rules.sonarsource.com}}.
In this format, information about the error should be provided in a concise language, as well as code examples of how to fix the error should be provided. 
The depicted security notification refers to a misuse of the Bouncy Castle Crypto API for asymmetric ElGamal encryption.

The aforementioned notification format has been integrated into the prototypical implementation of CRITICALMATE's analysis engine and IDE plugins (cf. Section~\ref{sec:immediate-feedback}).
In the following, we discuss potential improvements.

The participants of the study conducted by Tahaei et al.~\cite{tahaei2021security} imply that the most helpful part were the provided code examples in the notifications.
On the other side, they indicated that metadata was not relevant for them.
However, code examples can lead to the problem of mismatched examples.
One solution to tackle this problem would be to create parameterized code example templates that can be dynamically completed to match the specific scanned code snippet and the context of the software project.

This approach could also be taken further by automatically suggesting quick fixes for a given notification.
Developers can then choose to apply these fixes directly in their IDE.
In some cases, where a concrete fix might not be available, a snippet with placeholders for the developer to fill in could be suggested instead.
This would help developers in fixing security issues in their software more quickly and more accurately.
Some existing static analysis tools, such as Clippy\footnote{\url{https://github.com/rust-lang/rust-clippy}} or rust-analyzer\footnote{\url{https://rust-analyzer.github.io/}} from the Rust programming language ecosystem, already support such features for some problems.

Finally, the feedback of software developers to a notification, for example, the reason a notification was not helpful, can also be used to learn about the helpfulness of notifications and adapt future notifications' content.

\section{Related Work}\label{sec:related-work}
This section presents related work to SAST tools, which provide an IDE integration and consequently immediate feedback. 

Eclipse CogniCrypt\footnote{\url{https://www.eclipse.org/cognicrypt/documentation/codeanalysis}} is an open-source platform for static code analysis based on CrySL rules~\cite{kruger2019crysl}. Such rules can describe different error types indicating incorrect usage of cryptography libraries. For example, a \emph{constraint error} indicates that the static analysis detected an incorrect argument in a specific method call.

Find Security Bugs (FSB)\footnote{\url{https://find-sec-bugs.github.io}} is an open-source extension of the SpotBugs static analysis tool for security audits of web application in Java. It  can detect security issues, such as those described in the open web application security project (OWASP) Top 10.
Furthermore, it provides plugins for its integration in several IDEs, such as for Eclipse and IntelliJ.

The goals of our approach are very similar to both aforementioned works in respect to supporting software developers with immediate feedback within their IDEs. However, in addition, we aim for a user-centered approach that leverages developers feedback to build an adaptive system.

\section{Results and Conclusion}\label{sec:conclusion}
In this paper, we introduced an approach for supporting software developers to address possible security issues directly in their IDEs. Through continuous immediate feedback integrated in IDEs, developers can see notifications regarding security issues, such as misuse of cryptography libraries, as soon as code is typed in the code editor and the typed code has been analyzed by a SAST tool. Within the project \emph{Cybersecurity static analysis with immediate feedback} (CRITICALMATE), the University of Stuttgart and RIGS IT have jointly developed a prototype that integrates the analysis results of RIGS IT's security tool continuously into different IDEs, namely IntelliJ, Eclipse IDE, and Visual Studio Code. 
Preliminary measurements tell us that security notifications are shown to developers in average under a second. 
However, there are several factors that influence the response time from triggering the static code analysis until the visualization of the found security notifications.
To get more insights about these issues, we plan to conduct further experiments.

Furthermore, we plan to conduct a user study to get insights about the interaction of software developers with the immediate feedback and the adaptivity feature.
One possibility is to conduct a study that also conducts emotion recognition based on physiological signals or video input.
We can gain valuable feedback by recognizing confusion or frustration~\cite{Fernandez1998} of software developers while they are working with selected security notifications for the study.

One big challenge in the aforementioned approach is how to differentiate notifications that are actual false positives and what developers might label as false positives because they do not agree on the issue or perceive them as noise.
Furthermore, if there are many notifications regarding the same code snippet, these should be prioritized to be visualized sequentially based on, for example, their severity, to avoid cluttered visualization or missing important notifications.

\bibliographystyle{splncs04}
\bibliography{bibliography}

\end{document}